\documentclass[12pt]{article}
\usepackage{amsfonts,amsmath,amssymb,epsf,cite}

\usepackage{graphicx,color,subfigure}
\usepackage[usenames,dvipsnames]{pstricks}

\def\dee{\partial}

 \def\ep{{\epsilon}}

 \def\frac#1#2{{#1\over #2}}

\def\cD{{\cal D}}

\def\cG{{\cal G}}
\def\cC{{\cal C}}
\def\cJ{{\cal J}}

\def\nn{\nonumber}

 \def\ep{\epsilon}

 \def\ep{{\epsilon}}

 \def\frac#1#2{{#1\over #2}}

\def\cP{{\cal P}}
 
 \def\cO{{\cal O}}

 \def\ben{\begin{equation}}
\def\een{\end{equation}}
\def\beq{\begin{eqnarray}}
\def\eeq{\end{eqnarray}}
\def\bea{\begin{eqnarray}}
\def\eea{\end{eqnarray}}

\def\Psis{\Psi_s}
\def\bJ{{\bar{J}}}

\begin{document}
\begin{titlepage}
\thispagestyle{empty}
\begin{flushright}
UK/12-12\\
PUPT-2431\\
\end{flushright}

\bigskip

\begin{center}
\noindent{\Large \textbf
{Quantum Quench Across a Zero Temperature Holographic Superfluid Transition}}\\
\vspace{1.5cm} \noindent{
Pallab Basu$^{(a)}$\footnote{e-mail: pallab.basu@uky.edu},
Diptarka Das $^{(a)}$\footnote{e-mail: diptarka.das@uky.edu},
Sumit R. Das $^{(a)}$\footnote{e-mail: das@pa.uky.edu} and
Tatsuma Nishioka $^{(b)}$\footnote{e-mail: nishioka@princeton.edu}

\vspace{1cm}
 $^{(a)}$ {\it Department of Physics and Astronomy, \\
 University of Kentucky, Lexington, KY 40506, USA}\\
$^{(b)}${\it Department of Physics,\\
Princeton University, Princeton, NJ 08564, USA}}
\end{center}

\vspace{0.3cm}
\begin{abstract}

We study quantum quench in a holographic model of a zero temperature
insulator-superfluid transition. The model is a modification of that
of arXiv:0911.0962 and involves a self-coupled complex scalar field,
Einstein gravity with a negative cosmological constant, and Maxwell field
with one of the spatial directions compact. In a suitable regime of
parameters, the scalar field can be treated as a probe field whose
backreaction to both the metric and the gauge field can be ignored. We
show that when the chemical potential of the dual field theory lies
between two critical values, the equilibrium background geometry is a
AdS soliton with a constant gauge field, while the complex scalar
condenses leading to broken symmetry.  We then turn on a time
dependent source for the order parameter which interpolates between
constant 
values and crosses the order-disorder critical
point. In the critical region adiabaticity breaks down, but for a small
rate of change of the source $v$ there is a new small-$v$ expansion in
fractional powers of $v$. The resulting critical dynamics is dominated
by a zero mode of the bulk field. To lowest order in this small-$v$
expansion, the order parameter satisfies a time dependent
Landau-Ginsburg equation which has $z=2$, but non-dissipative. These
predictions are verified by explicit numerical solutions of the bulk
equations of motion.

\end{abstract}
\end{titlepage}
\newpage

\tableofcontents
\newpage

\section{Introduction and summary}

Recently there has been several efforts to understand the problem of
quantum or thermal quench \cite{sengupta,CCa,CCc}
in strongly coupled field theories using the
AdS/CFT correspondence \cite{Maldacena,GKP,W,AdSR}. This approach
has been used to explore two interesting issues. The first relates to
the question of thermalization. In this problem one typically
considers a coupling in the hamiltonian which varies appreciably with
time over some finite time interval. Starting with a nice initial
state (e.g. the vacuum) the question is whether the system evolves
into some steady state and whether this steady state resembles a
thermal state in a suitably defined sense. In the bulk description a
time dependent coupling of the boundary field theory is a time
dependent boundary condition. For example, with an initial AdS this
leads to black hole formation under suitable conditions. This 
is a holographic description of
thermalization, which has been widely studied over the past several
years \cite{janik,otherthermalization,holoentanglement} with
other initial conditions as well.  

Many interesting applications of
AdS/CFT duality involve a subset of bulk fields whose backreaction to
gravity can be ignored, so that they can be treated in a {\em probe
  approximation}. One set of examples concern probe branes in AdS
which lead to hypermultiplet fields in the original dual field
theory. Even though the background does not change in the leading
order, it turns out that thermalization of the hypermultiplet
sector is still visible - this manifests itself in the formation of
apparent horizons on the worldvolume \cite{dnt, otherapparent}.

The second issue relates to quench across critical points
\cite{sengupta,CCa,CCc}. Consider for example starting
in a gapped phase, with a parameter in the Hamiltonian varying slowly
compared to the initial gap, bringing the system close to a value of
the parameter where there would be an equilibrium critical point.  As
one comes close to this critical point, adiabaticity is
inevitably broken. Kibble and Zurek  \cite{kibblezurek, sengupta,gubsersondhi}
argued that in the critical region
the dynamics reflects universal features leading to scaling of various
quantities. These arguments are based on rather drastic
approximations, and for strongly coupled systems there is no
theoretical framework analogous to renormalization group 
which leads to such scaling.  For two-dimensional
theories which are {\em suddenly} quenched to a critical point,
powerful techniques of boundary conformal field theory have been used
in \cite{CCc} to show that ratios of relaxation times of one point
functions, as well as the length/time scales associated with the
behavior of two point functions of different operators, are given in
terms of ratios of their conformal dimensions at the critical point,
and hence universal.

In \cite{basudas} quench dynamics in the critical region of a finite chemical
potential 
holographic critical point was studied in a probe approximation. The
``phenomenological'' model used was that of \cite{liu1} which involves
a neutral scalar field with quartic self-coupling with a mass-squared
lying in the range $ -9/4 < m^2 < -3/2$ in the background of a {\em
  charged} $AdS_4$ black brane. The self coupling is large so that the
backreaction of the scalar dynamics on the background geometry can be
ignored. The background Maxwell field gives rise to a nonzero chemical
potential in the boundary field theory. In \cite{liu1} it was shown
that for low enough temperatures, this system undergoes a critical
phase transition at a mass $m_c^2$. For $m^2 < m_c^2$ the scalar field
condenses, in a manner similar to holographic superfluids
\cite{gubser,hhh,Basu,Ha,Basu2,cmtref1}. 
The
critical point at $m^2 = m_c^2$ is a standard mean field transition
at any non-zero temperature, and becomes a
Berezinski-Kosterlitz-Thouless transition at zero temperature, as in
several other examples of quantum critical transitions.  In
\cite{basudas} the critical point was probed by turning on a time
dependent source for the dual operator, with the mass kept exactly at
the critical value, i.e. a time dependent boundary value of one of the
modes of the bulk scalar. The source asymptotes to constant values at
early and late times, and crosses the critical point at zero source at
some intermediate time. The rate of time variation $v$ is slow
compared to the initial gap. As expected, adiabaticity fails as the
equilibrium critical point at vanishing source is approached. However,
it was shown that for any non-zero temperature and small enough $v$,
the bulk solution in the critical region can be expanded in {\em
  fractional} powers of $v$. To lowest order in this expansion, the
dynamics is dominated by a single mode - the zero mode of the
linearized bulk equation, which appears exactly at $m^2 = m_c^2$. The
resulting dynamics of this zero mode is in fact a {\em dissipative}
Landau-Ginsburg dynamics with a dynamical critical exponent $z=2$, and
the order parameter was shown to obey Kibble-Zurek type scaling.

The work of \cite{basudas} is at finite temperature - the dissipation in this model is of course due to the presence of a
black hole horizon and is expected at any finite temperature. It is
interesting to ask what happens at zero temperatures. It turns out
that the model of \cite{liu1} used in \cite{basudas} becomes subtle
at zero temperature. In this case, there is no conventional adiabatic
expansion even away from the critical point (though there is a
different low energy expansion, as in
\cite{fermisurface}). Furthermore, the susceptibility is finite at the
transition, indicating there is no zero mode. While it should be
possible to examine quantum quench in this model by numerical methods,
we have not been able to get much analytic insight.

In this paper we study a different model of a quantum critical point,
which is a variation of the model of insulator-superconductor
transition of \cite{Nishioka:2009zj}. The model of
\cite{Nishioka:2009zj} involves a {\em
  charged} scalar field minimally coupled to gravity with a negative
cosmological constant and a Maxwell field. One of the spatial
directions is compact with some radius $R$, and in addition one can
have a non-zero temperature $T$ and a non-zero chemical potential
$\mu$ corresponding to the boundary value of the Maxwell field. In the
absence of the scalar field this model has a line of Hawking-Page type
first order phase transitions in the $T$-$\mu$ plane which separates an
(hot) AdS soliton and a (charged) black brane. Exactly on the $T=0$
line, the two phases correspond to the AdS soliton with a constant
Maxwell scalar potential, and an extremal black hole. In
\cite{Nishioka:2009zj} it was shown that in the presence of a
minimally coupled charged scalar, the phase diagram changes. 
When the charge is large the scalar and the gauge fields can be
regarded as probe fields which do not affect the geometry.
Now
there is a phase with a trivial scalar and a phase with a scalar
condensate. In the boundary theory the latter is a superfluid
phase. This phase transition persists at zero temperature, where it
separates an unbroken phase at low chemical potential and a broken
phase - in both cases the background geometry is the AdS soliton,
while the gauge field is non-trivial in the superfluid phase. The phase diagram 
is given in Figure 9 of \cite{Nishioka:2009zj}.

The idea
now is to probe the dynamics of this insulator-superfluid transition at
zero temperature by turning on a time dependent source for the
operator dual to the charged field.
So long as the scalar is minimally coupled and the charge $q$ is large,
this would involve analyzing a coupled set of equations of the scalar
field and the gauge field.

 However, it turns out that a slight
modification of the model allows us to ignore the backreaction of the
scalar to the gauge field as well. This involves the introduction of a
quartic self coupling of the scalar $\lambda$.  Then in the regime
$\lambda \gg q^2 $ and $\lambda \gg \kappa^2$ (where $\kappa$ is the 
gravitational
coupling), we can consider the dynamics of the charged scalar in
isolation.

In this work we first show that in this regime of the parameters the
insulator-superfluid transition persists. Concretely, for a
sufficiently small negative $m^2$, there is a critical value of the
background chemical potential beyond which a nontrivial static
solution for the scalar becomes thermodynamically favored. Note that
unlike other models of holographic superconductors the trivial
solution does not become dynamically unstable. Rather the non-trivial
solution has lower energy. The
transition is a standard mean field critical transition.  The background
geometry remains an AdS soliton and the background gauge potential
remains a constant, which is the chemical potential $\mu$.  At the
transition, the linearized equation has a zero mode solution which is regular 
both at the boundary and at the tip.

We then
turn on a time dependent boundary condition and find that the
breakdown of adiabaticity for a small rate $v$ is characterized by
exponents which are appropriate for a dynamical critical exponent
$z=2$.  In a way quite similar to \cite{basudas} we find that in the
critical region there is a new small $v$ expansion in fractional
powers of $v$, and the dynamics is once again dominated by a zero
mode. The real and imaginary parts of the zero mode now satisfy a
coupled set of Landau-Ginsburg type equation with first order time
derivatives. However the resulting system is oscillatory rather than
dissipative - this is expected since the background geometry has no
horizon so that we have is a {\em closed} system.  The
order parameter is shown to obey a Kibble-Zurek type scaling. 
Finally we solve the bulk equations numerically and verify
the scaling property obtained from the above small-$v$ expansion.

Thermal quench in holographic superfluids with backreaction has been recently
studied in \cite{Bhaseen:2012gg}.
This work addresses a different issue - here the quench is applied to the system
in the ordered phase {\em away from the critical point} and the resulting late
time relaxation of the order parameter is studied. Our emphasis is on probing a
possible Kibble-Zurek scaling when the quench crosses the critical point.

In Section 2 we define the model and discuss its equilibrium phases. In Section 3 we study quantum quench in this model by turning on a time dependent source, discuss the breakdown of adiabaticity and show that the critical region dynamics is dominated by the zero mode, leading to scaling behavior. In Section 4 we present the results of a numerical solution of the equations, verifying the scaling behavior. In an appendix we discuss a Landau-Ginsburg model similar to the critical dynamics of our holographic model.

\section{The model and equilibrium phases}

The ``phenomenological'' holographic model we consider is a slight
variation of the model of \cite{Nishioka:2009zj}. The bulk action in
$(d+2)$-dimensions is
\ben  
S = \int d^{d+2}x {\sqrt{g}} \left[ \frac{1}{2\kappa^2} \left( R +
  \frac{d(d+1)}{L^2}\right) - \frac{1}{4}F_{\mu\nu}F^{\mu\nu} -
  \frac{1}{\lambda} \left( |\nabla_\mu \Phi - iqA_\mu \Phi|^2 - m^2
  |\Phi|^2 - \frac{1}{2} |\Phi|^4 \right) \right] \ ,
\label{1-1}
\een
where $\Phi$ is a complex scalar field and $A_\mu$ is an abelian gauge
field, and the other notations are standard. Henceforth we will use $L=1$
units.

One of the spatial
directions, which we will denote by $\theta$ will be considered to be
compact.
We will consider the regime
\ben
\lambda \gg q^2 \ ,~~~~~~\lambda \gg \kappa^2 \ .
\een
In this regime the scalar field is a probe field, and its backreaction
to both the metric and the gauge field can be ignored.

\subsection{The background}

The background
metric and the gauge field can be then obtained by solving the
Einstein-Maxwell equations with the appropriate periodicity condition
on $\theta$. It is well known that there are two possible solutions. The
first is the $AdS_{d+2}$ soliton,
\bea
ds^2 & = & \frac{dr^2}{r^2 f_{sl}(r)} + r^2 \left( -dt^2 +
\sum_{i=1}^{d-1} dx_i^2 \right) + r^2 f_{sl}(r) d\theta^2 \ ,\nn \\
f_{sl}(r) & = & 1 - \left( \frac{r_0}{r} \right)^{d+1} \ , \nn \\
A_t & = & \mu \ ,
\label{1-2}
\eea
with constant parameters $\mu$ and $r_0$. The periodicity of $\theta$ in
this solution is
\ben
\theta \sim \theta + \frac{4\pi}{(d+1)r_0} \ ,
\label{1-3}
\een
while the temperature can be arbitrary. The second solution is a
$AdS_{d+2}$ charged black hole
\bea
ds^2 & = & -r^2 f_{bh}(r) dt^2 + \frac{dr^2}{r^2 f_{bh}(r)}+r^2 \left(
\sum_{i=1}^{d-1} dx_i^2 + d\theta^2 \right) \ , \nn \\
f_{bh}(r) & = & 1 - \left[1+\frac{d-1}{2d}\left( \frac{\mu}{r_+} \right)^2
  \right] \left( \frac{r_+}{r} \right)^{d+1} + \frac{d-1}{2d} \left(
\frac{\mu}{r_+} \right)^2 \left( \frac{r_+}{r} \right)^{2d} \ , \nn \\
A_t & = & \mu \left[ 1 - \left( \frac{r_+}{r} \right)^{d-1} \right] \ .
\label{1-4}
\eea
The temperature of this black brane is 
\ben
T = \frac{r_+}{4\pi} \left[ d+1 - \frac{(d-1)^2}{2d} \left(
  \frac{\mu}{r_+} \right)^2 \right] \ ,
\label{1-5}
\een
while the period of $\theta$ is arbitrary. As shown in
\cite{Nishioka:2009zj}, this system undergoes a phase transition
between these two solutions when 
\ben
r_0^{d+1}= r_+^{d+1} \left[ 1 + \frac{d-1}{2d} \left(
    \frac{\mu}{r_+} \right)^2 \right] \ .
\label{1-6}
\een
The AdS soliton is stable when the temperature and the chemical
potential are small. At $T=0$ the transition happens at a critical
chemical potential $\mu_{c2}$ given by
\ben
\mu_{c2} = \frac{r_0 (d+1)
  (2d)^{\frac{d-1}{2(d+1)}}}{(d-1)^{\frac{d}{d+1}}(d+1)^{1/2}} \ .
\label{1-7}
\een

\subsection{Scalar condensate}

Consider now the scalar wave equation in
the AdS soliton background (\ref{1-2}).  We first rescale  
\ben
r \rightarrow \frac{r}{r_0}\ ,~~~~t \rightarrow tr_0 \ ,~~~~~\mu \rightarrow
\frac{\mu}{r_0} \ .
\label{1-7a}
\een
In the rest of the paper we will use these rescaled coordinates (i.e., $r_0=1$) and
chemical potentials.

For fields which depend only
on $t$ and $r$, the equation of motion is
given by 
\ben 
\left[ -\frac{1}{r^2}(\partial_t - i \mu)^2 + \frac{1}{r^d}
  \partial_r \left( r^{d+2} f_{sl}(r)\partial_r \right)\right]\Phi -
m^2 \Phi - \Phi |\Phi|^2 = 0 \ .
\label{1-8}
\een
In this paper we will consider $ -\frac{(d+1)^2}{4} < m^2 < -\frac{d(d-1)}{4}$. The
asymptotic behavior of the solution at the AdS boundary $r \rightarrow \infty$
is of
the standard form
\ben
	\Phi (r,t) = J(t)\, r^{-\Delta_-} [ 1+ O(1/r^2)]  + A(t)
        r^{-\Delta_+}[ 1+ O(1/r^2)] + \cdots \ ,
\label{1-9}
\een
where
\ben
	\Delta_\pm = \frac{d+1}{2} \pm \sqrt{m^2 + \frac{(d+1)^2}{4}}
        \ .
\label{1-10}
\een
In ``standard quantization'' $J(t)$ is the source, while the
expectation value of the dual operator is given by 
\ben
\langle \cO \rangle = A(t) \ .
\label{1-11}
\een
In ``alternative quantization'' the role of $J(t)$ and $A(t)$ are interchanged.
In this mass range both $\Delta_\pm$ are positive and both the
solutions of the linear equation vanish at the boundary. Thus the
nonlinear terms in the equation (\ref{1-8}) are subdominant - which is
why the leading solution near the boundary is the same as those of the
linear equation, as written above. 

We need to find time independent solutions of the equation
(\ref{1-9}).  Because of gauge invariance, we need to specify a gauge to
qualify what we mean by time independence. For the equilibrium
solution we require the solution to be {\em real} - this fixes the
gauge. Note that the tip of the soliton is locally two-dimensional
flat space. Therefore we need to require the solution to be regular at
the tip $r=1$.  This leads to the following boundary condition at $r
= 1$
\ben
\Phi(r) = \Phi_h + \Phi_h' (r-1) + \cdots \ ,
\label{1-12}
\een
where regularity requires
\ben
\Phi_h' = \frac{1}{(d+1)}\, \Phi_h (\Phi_h^2 + m^2) +
\frac{1}{(d+1)}\, 
\Phi_h \mu^2\ .
\label{1-13}
\een
To examine the phase structure we need to find time independent
solutions with a vanishing source. 

Clearly $\Phi = 0$ is always a solution. We have solved the equations
numerically and found that there is a
critical value of the chemical potential $\mu_{c1}$ beyond which there
is another solution with a non-trivial $r$ dependence which is
thermodynamically preferred. This means that for $\mu > \mu_{c1}$, the
operator dual to the bulk scalar has a vacuum expectation value,
i.e., the global $U(1)$ symmetry of the boundary theory is
spontaneously broken. Although this could happen both in the standard and
alternative quantizations, we need to check the critical value is less that that of the phase transition between the AdS soliton and AdS black hole: $\mu_{c1}<\mu_{c2}$. Otherwise, the scalar condensate phase is not available on the AdS soliton. 

\begin{figure}[htbp]
	\centering

	\includegraphics[width=9cm]{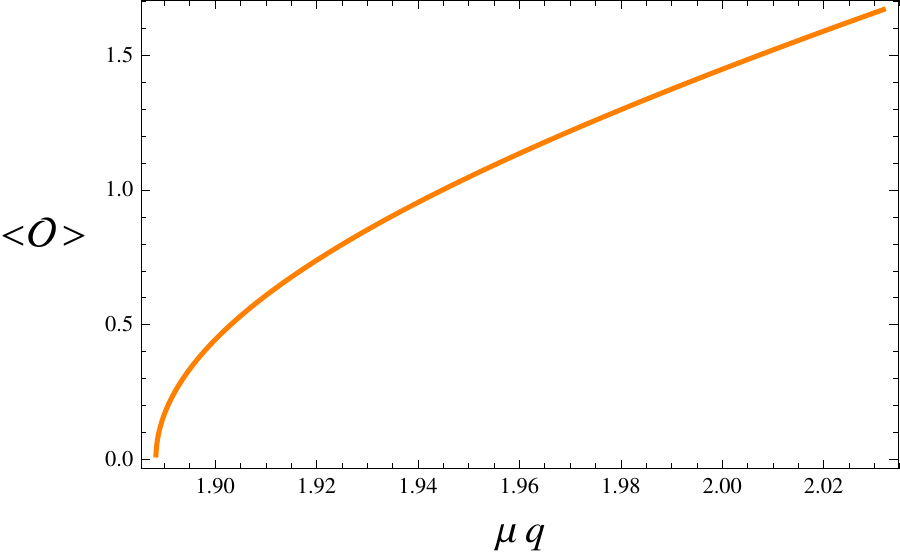}	
	\caption{The condensations of the scalar operators. }
	\label{fig:condensate}
\end{figure}

Figure \ref{fig:condensate} shows the behavior of
the expectation value $\langle \cO\rangle $ for $m^2 = -15/4$ for standard and
quantization.
We are plotting the condensation with respect to $\mu q$ and the phase transition happens at $\mu_{c1} q \sim 1.89$, which means the critical chemical potential is very small of order $O(1/q)$ in the probe limit. It follows from \eqref{1-7} that $\mu_{c1}$ is always much smaller than $\mu_{c2}\sim 1.86$ and there exists a scalar condensate phase on the AdS soliton.
Similarly for any given $m^2$, $\mu = \mu_{c1}$ is a critical point by letting $q$ be large enough.

This transition was first found in \cite{Nishioka:2009zj} for a {\em
  minimally coupled} complex scalar - in this case the backreaction to
the gauge field cannot be ignored, and the result follows from an
analysis of the coupled set of equations for the scalar and the gauge
field. In this case the gauge field introduces non-linearity in the
problem which is necessary for condensation of the scalar. What we
found is that a self-coupling does the same same job.

\subsection{The zero mode at the critical point}

To get some insight into this transition it is useful to write the
equation (\ref{1-8}) as a Schr{\"o}dinger problem. First define a new
coordinate 
\ben
{\rho(r)} = \int_r^\infty \frac{ds}{s^2 f_{sl}^{1/2}(s)} \ ,
\label{1-14}
\een
which is the ``tortoise coordinate'' for the AdS soliton. $\rho(r)$ is a
monotonic function of $r$ with the behavior
\bea
\rho & \sim&  1/r\ ,\qquad\qquad ~~~~~~~~~r \rightarrow \infty \ ,\nn \\
\rho & \rightarrow  & \rho_\star + \frac{2\sqrt{r-1}}{\sqrt{d+1}} \ ,~~~~~~~~ r \rightarrow 1 \ .
\label{1-14a}
\eea
For example, for $d=3$ (asymptotically $AdS_5$ spacetime)
soliton $\rho_\star = 1.311$. 
Let us now redefine the field by
\ben
\Phi(r,t) =
\frac{1}{[r(\rho)]^{\frac{(d-2)}{2}}}\left(
  \frac{d\rho}{dr}\right)^{1/2} \Psi (\rho,t) \ .
\label{1-15}
\een
Then $\Psi(\rho,t)$ satisfies the equation
\ben
\left[-\partial_t^2 +2i\mu \partial_t \right] \Psi = \cP \Psi - \mu^2
\Psi + 
\frac{r^{2-d}}{\sqrt{f_{sl}(r)}} |\Psi|^2 \Psi \ .
\label{1-16}
\een
The operator $\cP$ is
\bea
\cP & =& -\partial_\rho^2 + V_0(\rho)  \ ,\nn \\
V_0(\rho) & = & m^2 r^2 +
\frac{4d(d+2)r^{2d+2}-4d(d+3)r^{d+1}-(d-1)^2}{16r^{d-1}(r^{d+1}-1)}  \ ,
\label{1-17}
\eea
where 
$r$ has to be expressed as a function of $\rho$
using (\ref{1-14}). 

The potential $V_0(\rho)$ has the following behavior near the boundary
and the tip
\bea
V_0(\rho) & = & \frac{m^2+\frac{d(d+2)}{4}}{\rho^2} + O(\rho^2) \ ,\qquad\qquad \quad\rho
\rightarrow 0 \ ,  \nn\\
V_0(\rho) & = & -\frac{1}{4(\rho_\star - \rho)^2} + O(1)
\ ,\qquad\qquad \quad\rho
\rightarrow \rho_\star \ .
\label{1-18}
\eea
The behavior at the boundary $\rho = 0$ is of course the same as in
pure $AdS_{d+2}$.  The behavior near the tip $\rho = \rho_\star$ is in
fact the correct behavior expected from a flat two dimensional
space. Near the tip of the soliton the space becomes ${\mathbb R}^2 \times {\mathbb R}^{d-1}$ with
$y \equiv
(\rho_\star - \rho)$ playing the role of a radial variable and $\theta$
playing the role of the polar angle. Indeed with the redefined field 
\ben
{\tilde{\Psi}}(y) =\frac{ \Psi (\rho)}{ \sqrt{\rho_\star - \rho}}  \ ,
\label{1-19}
\een
the operator $\cP$ becomes, near $y = 0$, 
 the zero angular momentum Laplacian in two dimensions
\ben
\cP \,\underset{y \rightarrow 0}{\rightarrow}\, -
\frac{1}{y}\partial_y(y \partial_y) = -(\nabla_2)^2|_0  + {\rm{constant}}\ .
\label{1-20}
\een
In fact the eigenvalues of the operator $\cP$ which acts on $\tilde{\Psi}$ 
\ben
\cP = -(\nabla_2)^2|_{0} + V_1(y) \ ,~~~~~~\left( V_1(y) \equiv V_0(y) +\frac{1}{4y^2} \right) \ ,
\een
are all positive. For $d=3$ the proof is the following.
Let us rewrite the potential $V_1(y)$ as follows
\ben
V_1(y) = (m^2 + \frac{15}{4}) r^2 + V_2(y) 
\label{1-2-1}
\een
where
\ben
V_2(y) = \frac{1}{4} \left[
  \frac{1+3r^4}{r^2-r^6}+\frac{1}{y(r)^2} \right] 
\label{1-2-2}
\een
The term $V_2(y)$ is explicitly positive for all $r$. This may be seen as
follows. The condition for positivity of $V_2(y)$ is
\ben
\sqrt{\frac{r^6-r^2}{1+3r^4}} - y(r) \geq 0
\een
The inequality is saturated for $y=0$ ($r=1$). Furthermore the first
derivative of the left hand side becomes
\ben
\frac{1}{\sqrt{r^4-1}} \left[ \frac{3r^8+6r^4-1}{(1+3r^4)^{3/2}}
  -1 \right] 
\een
This can be explicitly checked to be positive for all $r>1$ (e.g. by
squaring the expression). Therefore $V_2(y) \geq 0$ for all $r>1$.
The first term in $V_1(y)$ in (\ref{1-2-1}) is the asymptotic
potential in $AdS_5$ - when $m^2 + \frac{15}{4} > -\frac{1}{4}$ (which
is the BF bound), this potential does not have any bound state. Since
$V_2(y)$ differs from this asymptotic potential by a positive
function, the full potential $V_1(y)$ does not have any bound state.

To look for a condensate in standard quantization, we need to find
time independent solutions of the equation (\ref{1-16}) which
satisfy the
boundary condition $J=0$ at $\rho = 0$ and is regular at the tip $\rho
= \rho_\star$. 
With these boundary conditions the operator $\cP$ has a
{\em discrete} and {\em positive} spectrum. This means that for
sufficiently large $\mu$ the operator
\ben
\cD \equiv \cP - \mu^2 \ ,
\label{1-21}
\een
will have a negative eigenvalue. This is what we found
numerically.

At the critical value $\mu = \mu_{c1}$ the operator $\cD$ has a zero
eigenvalue, i.e. a zero mode which satisfies the appropriate boundary
conditions both at the tip and at the boundary. This zero mode will play a
key role in the following.

Note that even though the operator $\cD$ has negative eigenvalues in
the condensed phase, the trivial solution does not become
unstable. This is clear from (\ref{1-16}) and from the fact the
spectrum of $\cP$ is positive, which shows that the frequencies of the
solutions to the linearized equation are all real.

Following the arguments of \cite{liu1} it can be easily checked that
the transition is standard mean field. This means that
\bea
\langle \cO \rangle_{J=0} & \sim & \sqrt{|\mu_{c1}-\mu|} \ , \nn \\
\langle \cO \rangle_{\mu = \mu_{c1}} & \sim & |J|^{1/3} \ .
\label{1-22}
\eea
We expect that this transition
extends to non-zero temperature, though we have not checked this
explicitly.

\section{Quantum quench with a time dependent source}

We will now probe the quantum critical point by quantum quench with a
time dependent homogeneous source $J(t)$ for the dual operator
$\cO$, with the chemical potential tuned to $\mu = \mu_{c1}$.  The
function $J(t)$ will be chosen to asymptote to constants at early and
late times, e.g.
\ben
J(t) = J_0 \tanh (vt) \ .
\label{2-1}
\een
Note that we are using units with $r_0=1$. The system then crosses the
equilibrium critical point at time $t=0$.
The idea is to start at
some early time with initial conditions provided by the {\em
  instantaneous solution} and calculate the one point function $\langle \cO (t)\rangle $.
In standard quantization this means that we impose a time dependent
boundary condition as in (\ref{1-9}) and calculate $A(t)$.
In alternative quantization the
source should equal $A(t)$. In this paper we discuss the problem in
standard quantization : the treatment in alternative quantization is
similar.

\subsection{Breakdown of adiabaticity}

With a $J(t)$ of the form described above (e.g. (\ref{2-1})), one
would expect that the initial time evolution is adiabatic for small
$v$ so long as $J_0$ is not too small. As one approaches $t = 0$
adiabaticity inevitably breaks down and the system gets excited.  In
this subsection we determine the manner in which this happens. 

An adiabatic solution of (\ref{1-16}) is of the form
\ben
\Psi(\rho,t) = \Psi^{(0)}(\rho,J(t))+ \epsilon \Psi^{(1)}(\rho,t) +
\epsilon^2 \Psi^{(2)} + \cdots \ , 
\label{2-6}
\een
where $\epsilon$ is the adiabaticity parameter. The leading term is the
instantaneous solution of (\ref{1-16}), which is (using the definition
(\ref{1-21}))
\ben
\cD \Psi^{(0)} + G(\rho) |\Psi^{(0)}|^2 \Psi^{(0)} = 0 \ ,
\label{2-7}
\een
satisfying the required boundary condition. 
Here we have defined 
\ben
G(\rho) \equiv \frac{r^{2-d}}{\sqrt{f_{sl}(r)}} \ .
\label{2-4}
\een
From (\ref{1-22}) we know
that for a real $J(t)$, this is real and has a form
\ben
\Psi^{(0)} \sim \rho^\alpha J(t) \left[ 1 + O(\rho^2) \right] +
\rho^{1-\alpha} \left[ J(t) \right]^{1/3} \left[ 1 + O(\rho^2) 
\right] \ ,
\label{2-8}
\een 
where
\ben 
\alpha \equiv \Delta_- - d/2 \ .
\label{2-8b}
\een
This follows from the equations (\ref{1-14a}), (\ref{1-15}) and  (\ref{1-9}).
The adiabatic expansion now proceeds by replacing $\partial_t
\rightarrow \epsilon \partial_t$ in (\ref{1-16}) substituting
(\ref{2-6}) and equating terms order by order in $\epsilon$. The
$n$-th order contribution $\Psi^{(n)}$ satisfies a {\em linear,
  inhomogeneous} ordinary differential equation with a source term
which depends on the previous order solution $\Psi^{(n-1)}$. To lowest
order we have the following equations for the real and imaginary parts
of $\Psi^{(1)}$
\bea
\left[ \cD + 3 G(\rho) (\Psi^{(0)})^2 \right] ({\rm Re}\, \Psi^{(1)}) &
= & 0  \ ,\nn \\
\left[ \cD +  G(\rho) (\Psi^{(0)})^2 \right] ({\rm Im}\, \Psi^{(1)}) &
= & 2\mu\, \partial_t \Psi^{(0)} \ .
\label{2-9}
\eea
Note that in these equations the time dependence of $J(t)$ should be ignored.
The full function $\Psi$ must satisfy the boundary condition $\underset{\rho \rightarrow 0}{\rm
  lim} \left[ \rho^{-\alpha} \Psi (\rho,t)
  \right] = J(t) $. This means that the adiabatic corrections must
start with the subleading terms, $\Psi^{(1)} \sim \rho^{1-\alpha}$ as
$\rho \rightarrow 0$ and has to be regular as $\rho \rightarrow
\rho_\star$. These provide the boundary conditions for solving the
equations (\ref{2-9}). Consider first the equation for ${\rm Im}\,
\Psi^{(1)}$. Since the time dependence of $\Psi^{(0)}$ is entirely
through $J(t)$ the solution may be written as
\ben
{\rm Im}\,
\Psi^{(1)} (\rho,t) =2\mu\, {\dot{J}(t)}\int_0^{\rho_\star} d\rho^\prime\, \cG (\rho,
\rho^\prime)\, \frac{\partial \Psi^{(0)}}{\partial J(t)}
(\rho^\prime,J(t)) \ ,
\label{2-10}
\een
where $ \cG (\rho,\rho^\prime)$ is the Green's function for the
operator $ \cD +  G(\rho) (\Psi^{(0)})^2 $, 
\begin{align}
\cG(\rho,\rho^\prime) = \frac{1}{W(\psi_1,\psi_2)} \, \psi_1(\rho^\prime)
\psi_2(\rho) \ , \quad \rho < \rho^\prime \ ,\\
	     =  \frac{1}{W(\psi_1,\psi_2)} \, \psi_2(\rho^\prime)
             \psi_1(\rho)\ , \quad \rho > \rho^\prime,
\label{2-11}
\end{align}
where $\psi_1$ and $\psi_2$ are solutions of the homogeneous equation 
$\left[ \cD +  G(\rho) (\Psi^{(0)})^2 \right] \psi_{1,2} = 0 $ which
satisfy the appropriate boundary conditions at the tip $\rho =
\rho_\star$ and at the boundary $\rho = 0$ respectively. The Wronskian
$W(\psi_1,\psi_2)$ for this operator is clearly constant and is
conveniently evaluated near the tip. Near $\rho = \rho_\star$ these
solutions behave as
\ben
\psi_1 \sim C\sqrt{\rho_\star-\rho}\ ,~~~~~~~~\psi_2 \sim
A\sqrt{\rho_\star-\rho} + B\sqrt{\rho_\star-\rho}~ \log (\rho_\star -
\rho) \ ,
\label{2-12}
\een
where $A,B,C$ are constants which depends on $J(t)$ \footnote{Note that in the
equations 
(\ref{2-9}) the time is simply a parameter.}. Thus the Wronskian is 
\ben
W(\psi_1,\psi_2) = -BC \ .
\label{2-13}
\een
As noted in the previous section, the operator $\cD$ has a zero
mode, i.e. $\left[ \cD +  G(\rho)
  (\Psi^{(0)})^2 \right]$ has a zero mode when $\Psi^{(0)}=0$,
i.e. exactly at the equilibrium critical point. Thus, at this point we
must have $B=0$. This is why the first
adiabatic correction $  {\rm Im}\,
\Psi^{(1)} (\rho,t) $ diverges at this point. For small $J(t)$ we can use
perturbation theory to
estimate the value of $B$.  For small $J$ the zeroth order solution
$\Psi^{(0)}$ behaves as $J^{1/3}$ (the first term in (\ref{2-8}) is
subdominant). 
This is explicit to all orders in the expansion of the solution around
the boundary. However, this is also justified by the results of the
next section where we show that in the critical region the dynamics is
dominated by a zero mode. The coefficient of the zero mode can be seen
to be proportional to $J^{1/3}$ using a regularity argument similar to
that in \cite{liu1} so that the additional term in the operator
  behaves as
$G(\rho)  (\Psi^{(0)})^2  \sim [J(t)]^{2/3}$. This yields $B \sim
J^{2/3}$ as well. Thus the Green's function which appears in
(\ref{2-10}) behaves as $J^{-2/3}$ so that the correction behaves as 
\ben
{\rm Im}\,\Psi^{(1)} \sim \frac{{\dot{J}(t)}}{J^{2/3}}  \frac{\partial
  \Psi^{(0)}}{\partial J(t)}  \sim \frac{{\dot{J}(t)}}{J^{4/3}} \ .
\label{2-14}
\een
The same argument shows that ${\rm Re}\,\Psi^{(1)} = 0$, so that 
 $| \Psi^{(1)} | \sim \frac{{\dot{J}}}{J^{4/3}}$ as well. Therefore
adiabaticity breaks when 
\ben
| \Psi^{(1)} | \sim | \Psi^{(0)} | \implies {\dot{J}(t)} \sim J^{5/3} \ .
\label{2-15}
\een
For sources which vanish linearly at $t = 0$, i.e. $J(t) \sim vt$  (e.g. of the
form
(\ref{2-1})) this means that if the source is turned on at some early
time, adiabaticity breaks at a time
\ben
t_{adia} \sim v^{-2/5} \ .
\label{2-16}
\een
while at this time the value of the order parameter $\langle \cO\rangle$ is
\ben
\langle \cO (t_{adia})\rangle  \sim [J(t_{adia})]^{1/3} = [v t_{adia}]^{1/3} \sim
v^{1/5} \ .
\label{2-17}
\een
With the usual adiabatic-diabatic assumption, these 
exponents lead to Kibble-Zurek scaling for a dynamical critical
exponent $z=2$, even though the underlying dynamics is relativistic
and non-dissipative. From the above analysis it is clear that this
happened because the leading adiabatic correction is provided by the
chemical potential term, which multiplies a first order time
derivative of the bulk field.

\subsection{Critical dynamics of the order parameter}

The breakdown of adiabaticity means that an expansion in time
derivatives fail. In this subsection we show, following closely the
treatment of \cite{basudas}, that we now have a {\em different} small
$v$ expansion in {\em fractional} powers of $v$ during the period
when the sources passes through zero. This will lead to a scaling form
of the order parameter in the critical region.
In the following we will
demonstrate this for the case where $J(t) \sim vt$ near $t \approx
0$. However the treatment can be easily generalized to a $J(t) \sim
(vt)^n$ for any integer $n$.

To establish this, it is convenient to 
separate out the source term in the field $\Psi
(\rho,t)$,
\ben
\Psi (\rho,t) = \rho^{\alpha}J(t) + \Psis(\rho,t) \ , ~~~~~~\alpha =
\Delta_- - d/2 \ ,
\label {2-2}
\een
where we have used the relation (\ref{1-15}) and the fact that near
the boundary $\rho \sim 1/r$.
The equation of motion (\ref{1-16}) then becomes
\bea
-\partial_t^2 \Psis + 2i\mu\, \partial_t \Psis & = & (\cD
\rho^\alpha)J(t) + \cD \Psis + G(\rho) \left[ \rho^{3\alpha}[J(t)]^3
  + \rho^{2\alpha}[J(t)]^2
(2\Psis + \Psis^\star) \right] \nn \\
& & + G(\rho) \left[ \rho^\alpha J(t) (2 |\Psis|^2 + \Psis^2) +
|\Psis|^2 \Psis \right] \nn \\
& & +\rho^\alpha [\partial_t^2 J - 2i\mu \partial_t J] \ .
\label{2-3}
\eea
This separation is useful because we know that in the presence of a
constant source $J(t) = \bJ$, the static solution has the asymptotic form
\ben
\Psis \sim \rho^{1-\alpha} \left[ |\bJ|^{1/3} + O(\rho^2) \right] + \bJ
\rho^{\alpha+2}\left[  1 + O(\rho^2) \right] \ ,
\label{2-5}
\een
which follows from (\ref{1-22}). 

The scaling relations (\ref{2-16}) and (\ref{2-17}) suggest that we
perform the following rescaling of the time and the field
\ben
t = v^{-2/5} \eta \ ,~~~~~~~~~~~~~~\Psi_s = v^{1/5} \chi \ .
\label{2-18}
\een
In the critical region we can now use $J(t) = vt = v^{3/5}\eta$ and rewrite
(\ref{2-3}) as an expansion in powers of $v^{2/5}$,
\ben
\cD \chi =  v^{2/5} \left[2i\mu \partial_\eta \chi - G(\rho)
  |\chi|^2 \chi - \eta (\cD \rho^\alpha) \right] +O(v^{4/5}) \ .
\label{2-19}
\een 
As noted above, because of the boundary condition at $\rho =0$ and the
regularity condition at $\rho = \rho_\star$ the spectrum of $\cD$ is discrete. 
Let $\varphi_n$ be the orthonormal set of eigenfunctions of the operator $\cD$
\ben
\cD \varphi_n (\rho) = \lambda_n \varphi_n (\rho) \ ,~~~~~n=0,1,\cdots \ ,
\label{2-20}
\een
with $\lambda_0 = 0$.
$\varphi_0(\rho)$ is the zero
mode which we discussed earlier. Since $\mu$ has been tuned to be equal to
$\mu_{c1}$, all the higher eigenvalues are positive.

We now expand
\ben
\chi(\rho,\eta) = \sum_n \chi_n (\eta) \varphi_n(\rho) \ ,
\label{2-21}
\een
and rewrite the equation (\ref{2-19}) in terms of the modes $\chi_n
(\eta)$
\ben
\lambda_n \chi_n = v^{2/5} \left[ 2i\mu (\partial_\eta \chi_n) -
  \sum_{n_1n_2n_3} \cC^{n}_{n_1n_2n_3} \chi_{n_3}^\star \chi_{n_2}
  \chi_{n_1} + \cJ_n \eta \right] + O(v^{4/5}) \ ,
\label{2-22}
\een
where we have defined
\bea
\cJ_n & = & \int d\rho \varphi_n^\star (\rho) (\cD \rho^\alpha) \ , \nn \\
\cC^{n}_{n_1n_2n_3} & = & \int d\rho \varphi_n^\star(\rho)
\varphi_{n_3}^\star(\rho) 
\varphi_{n_2}(\rho) \varphi_{n_1}(\rho) G(\rho) \ .
\label{2-23}
\eea
It is clear from (\ref{2-22}) that the zero mode part of the bulk
field dominates the dynamics in the critical region.
In fact for small $v$ a solution is of the form
\ben
\chi_n(\eta) = \delta_{n0} \xi_0(\eta) + v^{2/5} \xi_n + O(v^{4/5}) \ .
\label{2-24}
\een
The zero mode satisfies a $z=2$ Landau-Ginsburg equation 
\ben
-2i\mu \partial_\eta \xi_0 + \cC^0_{000} |\xi_0|^2 \xi_0 +\cJ_0 \eta =
0 \ .
\label{2-25}
\een
Reverting back to the original variables we therefore have
\ben
\Psi_s (\rho,t,v) = v^{1/5} \Psi_s (\rho,t v^{2/5},1) \ ,
\label{2-26}
\een
which implies a Kibble-Zurek scaling for the order parameter with $z=2$
\ben
\langle \cO(t,v)\rangle  = v^{1/5} \langle \cO (v^{2/5}t,1)\rangle   \ .
\label{2-27}
\een
Note that the effective Landau-Ginsburg equation (\ref{2-25}) is not dissipative because the first order time derivative is multiplied by a purely imaginary constant. In fact, in the absence of a source term the quantity $\frac{1}{2} (|\xi_0|^2)^2$ is independent of time. 

Beyond the critical region, we cannot use the approximation $J(t) \sim vt$ and there is no useful simplification in terms of the zero mode. However, the boundary conditions at the tip are perfectly reflecting boundary conditions (as appropriate for the origin of polar coordinates in two dimensions) so that there is a conserved energy in the problem. This is in contrast to a black hole background where there is a net ingoing flux at the horizon causing the system to be dissipative. Indeed in the quench problem considered in \cite{basudas} arguments similar to those used in this section also led to an effective Landau-Ginsburg dynamics with $z=2$, but which is dissipative.

In the appendix we analyze a Landau-Ginsburg toy model motivated by the results of this section.

\section{Numerical results}

In this section we summarize our numerical results. We have solved the bulk equation of motion numerically for $d=3$. The results for different values of $m^2$ are similar. We present detailed results for $m^2 = -15/4$. In this case the critical value of the chemical potential is $\mu_{c1} q\approx 1.88$.

We discretize the partial
differential equations (PDEs) \eqref{1-16} 
(written in the $y$-coordinate) in a radial
Chebyshev grid to study the numerical problem.  Once discretized in
radial direction, the PDEs become a series of ordinary differential equations
(ODEs) in the temporal variable. The resulting ODEs are solved with a standard ODE solver
(e.g. CVODE). The time dependence is chosen to be of the form
dependent source as in \eqref{2-1}. 
In principle one may study with any kind of time dependent source.

We will consider the problem in two regimes.
The first is the ``slow" regime where we expect our analytic arguments to be accurate, 
the other is a ``fast" regime where there is no adiabatic region whatsoever. In the slow regime we will try to zoom on the scaling region around the phase transition. 
In the fast
regime we will find a large deviation from the adiabatic behavior and possible
chaotic behavior.  

\subsection{Slow regime}

\begin{figure}[h!]
\begin{center}
 \includegraphics[scale=0.7]{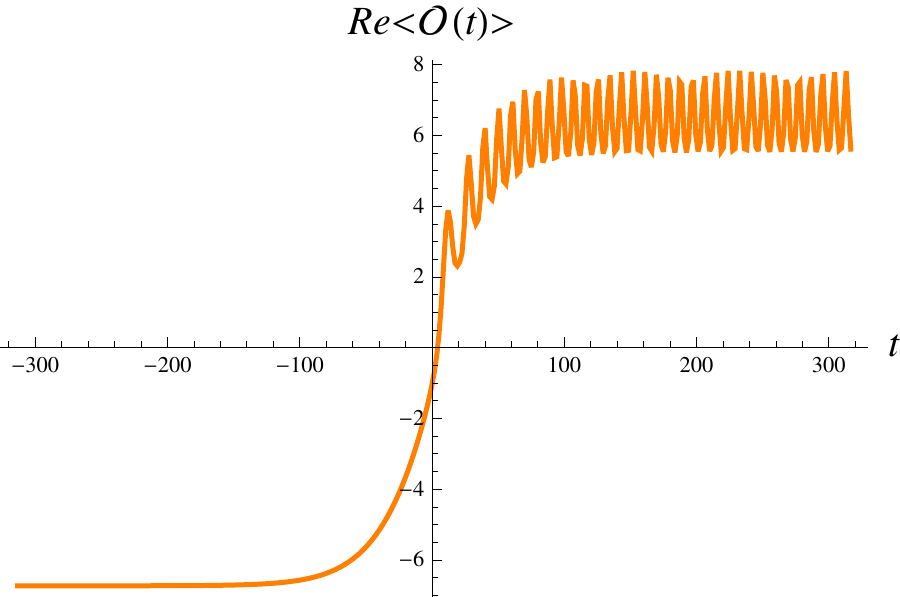}
\end{center}
\caption{The plot of ${\rm Re}\, \langle\cO (t)\rangle $ with $v=0.02$.}
\label{fig:quench}
 \end{figure}

Since our main interest is quench through the critical point, we concentrate mainly near the phase transition. We choose 
$\mu q=\mu_{c1}q\,(\approx 1.88)$, so that the system is critical in the absence of any source.
In the presence of a time dependent source of the form (\ref{2-1}) we calculate the bulk field ${\tilde{\Psi}(t)}$ and extract from this the value of $\langle \cO(t)\rangle$ of the dual field theory. A typical plot of the real part of $\langle \cO(t)\rangle$ for slow quench through the phase transition is presented in Figure \ref{fig:quench}.

Clearly the late time behavior is oscillatory, reflecting the fact that we are dealing with a closed and non-dissipative system.

We then zoom on the critical
region near $t = 0$ for various value of $v$ to look for any scaling behavior. One way to look for this is to consider the behavior of $\langle \cO\rangle $ at $t=0$. Equation (\ref{2-27}) then predicts a scaling behavior $\langle \cO (0)\rangle  \sim v^{1/5}$.

\begin{figure}[h!]
\begin{center}
 \includegraphics[scale=0.8
]{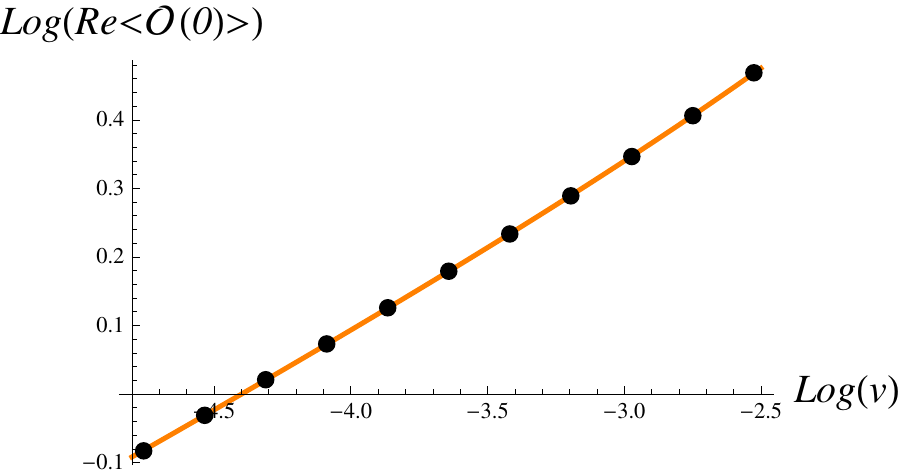}
\end{center}
\caption{The plot of $\log({\rm Re}\,\langle \cO(0)\rangle )$ vs $\log(v)$. We also plotted the closest fit (see text).}
\label{fig:quenchO}
 \end{figure}

Figure \ref{fig:quenchO} shows a plot of $\log({\rm Re}\, \langle \cO (0)\rangle )$ for different $v$.
We fit the data points with a function $f(x)=A+B x+C/x$, where $x$ is $\log(v)$. Here we kept a sublinear ($O(1/x)$) term to understand how the fit function
 approaches a linear regime. From our analytic argument we expect $B=1/5$. A fit of the numerical results yields
$f(x)=0.794- 0.490 /x + 0.206 x$. Changing the number of fit points and range changes the values of fit parameters a bit, however we always
get a value of $B$ which is close to $1/5$ with only a few percentage deviation. The imaginary part $( {\rm Im}\,\langle \cO  (0) \rangle)$ also satisfies the same scaling.

\subsection{Fast regime}

\begin{figure}[h!]
 \begin{center}
 \includegraphics[scale=0.7]{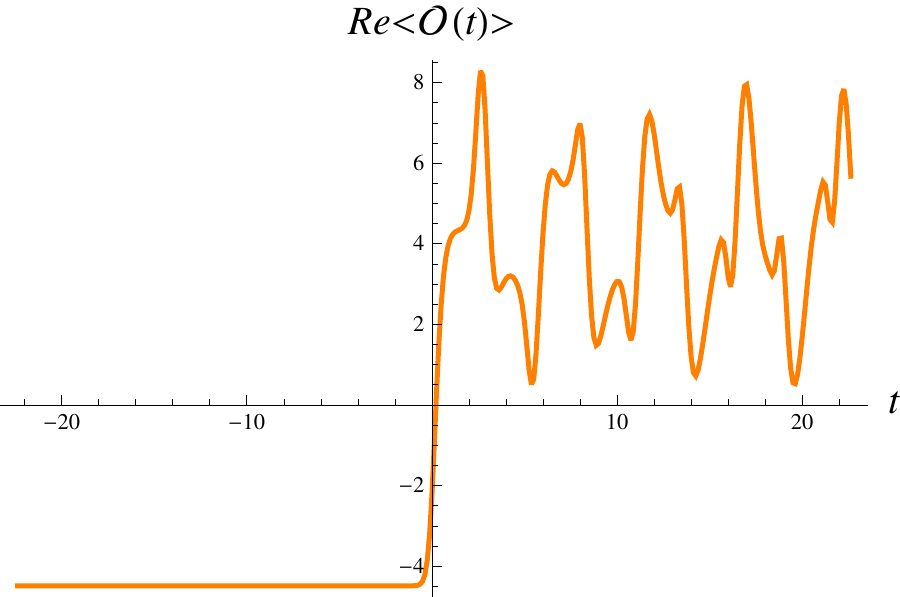}
 \end{center}
\caption{Plot of Re\,$\langle \cO (t)\rangle $ showing chaotic behavior with a large value of $v=2.0$ and $\mu=1$.}
\label{fig:quench2}
\end{figure}

In the fast regime we see a large deviation from the adiabatic behavior, as shown in 
Figure \ref{fig:quench2}.  

The motion in this regime becomes possibly chaotic. Here we have a system with a conserved energy. Once we put some energy in the system, the non-linearity possibly takes the system over the whole phase space (Arnold diffusion).  It is expected that if we wait long enough the probe approximation actually breaks down \cite{Dias:2012tq} and we have to consider the fully backreacted problem. We plan to attack this problem in the near future.

\section*{Acknowledgements}

This work is partially supported by National Science Foundation grants
PHY-0970069, PHY-0855614 and PHY-1214341. The work of T.N. was supported in part by the US NSF under Grant No.\,PHY-0756966. Some of these results were presented in the
workshop  ``Holographic Thermalization" at Lorentz Center, University of Leiden,
October 8-13, 2012. S.R.D. would like to thank M.\,Bhaseen and other participants
of the workshop for discussions, and the Lorentz Center for hospitality.
T.N. is grateful to K.\,Schalm for discussions.

\appendix
\section{Adiabatic and scaling analysis of a toy model}

In this appendix, we consider a $(0+1)$-dimensional toy model which follows the equation
\beq
\- 2i\mu \dot{\phi}+(m^2-\mu^2)\phi + \phi|\phi|^2 = J(t) \ .
\label{toy}
\eeq
The function $J(t)$ asymptotes to constants at early and late times and passes through zero in a linear fashion at some intermediate time, e.g.
\ben
J(t) = J_0 \tanh (vt) \ .
\een

\subsection{Adiabaticity}
We first derive conditions for breakdown of adiabaticity near the critical point $m^2 = \mu^2$ and $J(t) = 0$. We carry
out adiabatic expansion as following: 
\beq
\dee_t \rightarrow \ep \dee_t\ , \quad \phi \rightarrow \phi_0 (J(t))+ \ep \phi_1 (t)+ \cdots \ ,
\eeq
where $\phi_0(J(t))$ is the (real) adiabatic solution given by
\ben
\phi_0(J(t)) = [J(t)]^{1/3} \ .
\een
The solution to  $O(\ep^2)$ is ,
\beq
\phi = \phi_0 [J(t)]+ \ep\, i\, \frac{2\mu}{\phi_0^2}\dot{\phi_0}  + \ep^2 
\frac{1}{3\phi_0^2}\bigg(8\mu^2(\frac{\dot{\phi_0}^2}{\phi_0^3}-\frac{\ddot{
\phi_0}}{2\phi_0^2})- \frac{4\mu^2}{\phi_0^3}\dot{\phi_0}^2 \bigg) + O(\epsilon^3) \ .
\eeq
where
The breakdown of adiabaticity happens when,
\bea
\frac{2\mu}{\phi_0^2}\dot{\phi_0} &\sim& \phi_0  \ ,\\
\frac{1}{3\phi_0^2}\bigg(8\mu^2(\frac{\dot{\phi_0}^2}{\phi_0^3}-\frac{\ddot{
\phi_0}}{2\phi_0^2})- \frac{4\mu^2}{\phi_0^3}\dot{\phi_0}^2 \bigg) &\sim& \phi_0 \ ,
\eea
For $J(t) = \tanh(vt) \sim vt$ the above two equations translate into,
\beq
\mu &\sim& t^{5/3}  v^{2/3} \ .
\eeq
Thus if $\mu$ is of $O(1)$ then the above equations give us,
\beq
t &\sim& v^{2/5}  \ .
\eeq

\subsection{Scaling behavior}
Now sitting at the critical point we study the behavior of the scaling solution
with $\mu = O(1)$. From the adiabatic analysis we expect scaled time,
$\bar{t} = v^{2/5}t$. \\ We write the field $\phi$ as $\chi + i \xi$ and the source
as $J_\text{R} + i J_\text{Im}$, where both $J_\text{R}$ and $J_\text{Im}$ go as $v t$. To find the scaling exponents we extract the $v$ dependencies as,
\ben
t = v^\alpha \bar{t} \ ,\qquad \chi = v^\beta \bar{\chi} \ ,\qquad 
\xi = v^\gamma \bar{\xi} \ .
\een
Consistency of the equations demand,
\ben
\alpha = -\frac{2}{5} \ ,\quad 
\beta =\frac{1}{5} \ ,\quad 
\gamma = \frac{1}{5} \ .
\een
This determines the scaling behavior of the field $\phi$ at $m^2 = \mu^2$ and with $\mu$ of
$O(1)$ as,
\beq
\phi(t, v) = v^{1/5} \phi(v^{2/5}t,1) \ .
\eeq
This agrees with our expectation from adiabatic analysis, and also has been 
confirmed numerically in Section 4.

\subsection{Late time behavior} 

At late times the source $J(t)$ can be treated to be a constant and the solution can be obtained by perturbing around the static solution, $\phi_{static} = J^{1/3}$. It is then straightforward to see that the solution $\phi(t)$ is oscillatory with a frequency $\omega = \frac{J^{2/3}}{\sqrt{2}\mu}$.

\end{document}